\documentclass[twocolumn,prl]{revtex4}
\usepackage{graphicx}
\usepackage{amsmath}

\def\eq#1{Eq.~(\ref{#1})}
\def\fig#1{Fig.~\ref{#1}}

\def\12{\frac{1}{2}}
\def\kt{k_BT}
\def\nm{\,\rm nm}
\def\pn{\,\rm pN}

\def\hz{\,\rm Hz}

\newcommand{\comment}[1]{}

\begin{document}

\title{Bistability of cell-matrix adhesions resulting from non-linear receptor-ligand dynamics}
\author{T.~Erdmann}
\author{U.~S.~Schwarz}

\affiliation{University of Heidelberg, Im Neuenheimer Feld 293, D-69120 Heidelberg, Germany}

\begin{abstract}
  Bistability is a major mechanism for cellular decision making and
  usually results from positive feedback in biochemical control
  systems.  Here we show theoretically that bistability between
  unbound and bound states of adhesion clusters results from positive
  feedback mediated by structural rather than biochemical processes,
  namely by receptor-ligand dissociation and association dynamics
  which depend non-linearly on mechanical force and receptor-ligand
  separation. For small cell-matrix adhesions, we find rapid switching
  between unbound and bound states, which in the initial stages of
  adhesion allows the cell to explore its environment through many
  transient adhesions.
\end{abstract}

\maketitle

In cell-matrix adhesion, cells use integrin-based contacts to collect
information about their environment \cite{c:geig02}. The integrin
receptors form two-dimensional clusters in the plasma membrane and are
stabilized by connections to the cytoskeleton. Their extracellular
domains reversibly bind ligands from the extracellular matrix like
fibronectin, which present the ligand epitope through a polymeric
tether. In order to efficiently explore a newly encountered surface and
to commit itself to adhesion only if an appropriate combination of local
signals is present, it is favorable for cells to start by establishing
many small and transient adhesions sites. In the following we introduce
a simple theoretical model which predicts that small cell-matrix
adhesions are characterized by bistability between bound and unbound
states due to well-established principles of receptor-ligand dynamics.
For biochemical control structures, positive feedback is well-known to
result in bistability and switch-like behavior, for example in the cell
cycle and in the MAPK-cascade \cite{s:ange04}. Here we show that
bistability in biological systems can also result from structural
processes.

\begin{figure}
\begin{center}
\includegraphics[width=0.8\columnwidth]{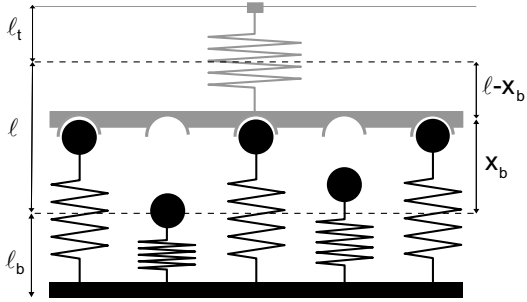}
\caption{Schematic representation of our model for an adhesion cluster
  with $N_t$ bonds, out of which $i$ are closed at a given time (here
  $N_t = 5$ and $i = 3$). The force transducer at the top and the
  ligand tethers at the bottom are modeled as harmonic springs, with
  rest lengths $\ell_t$ and $\ell_b$ and spring constants $k_t$ and
  $k_b$, respectively. The positions of the unloaded springs (dashed
  lines) are separated by the unloaded receptor-ligand distance
  $\ell$. The extensions of the loaded transducer and bond springs are
  denoted $x_t$ and $x_b = \ell - x_t$, respectively.}
\label{cartoon}
\end{center}
\end{figure}

\fig{cartoon} shows the model studied in the following. The mechanical
properties of the cell envelope holding the receptors and of the
polymeric ligands are represented by harmonic springs. We consider a
situation in which $N_t$ receptor-ligand pairs are arranged in parallel
along the cell-substrate interface. At a given time $t$, there are $i$
closed and $N_t-i$ open bonds. The dynamic variable $i$ ranges from $0$
(completely unbound state) to $N_t$ (completely bound state). The
probabilities for the $N_t+1$ states are denoted by $p_i(t)$ ($0 \le i
\le N_t$) and their time evolution is described by a one-step master
equation:
\begin{equation}\label{MasterEquation}
\frac{dp_i}{dt} = r_{i+1} p_{i+1} + g_{i-1} p_{i-1} - ( r_i + g_i ) p_i\,.
\end{equation}
This equation states that $i$ can either decrease due to rupture of a
closed bond (reverse rate $r_i$) or increase due to binding of an open
bond (forward rate $g_i$).

The reverse rate is $r_i = i k_{off}(i)$ because any of the $i$ closed
bonds can be the next to break. The single bond dissociation rate
increases exponentially with force according to the Bell equation,
$k_{off}(i) = k_0 e^{F_b(i)/F_0}$, with the unstressed dissociation rate
$k_0$ and the internal force scale $F_0$ \cite{c:bell78}. The force
$F_b(i)$ acting on each of the closed bonds changes dynamically and
follows from mechanical equilibrium for the arrangements of springs
depicted in \fig{cartoon}:
\begin{equation} \label{Force_i}
F_b(i) = \frac{k_b \ell}{1 + i (k_b/k_t)}\,.
\end{equation}
Here, $k_b$ represents the stiffness of the polymeric tethers carrying
the ligands and $k_t$ represents the effective stiffness of the cell
envelope. The distance $\ell$ is the average distance between
receptors and ligands if the cluster is completely unbound.
\eq{Force_i} shows that the more bonds are closed, the less force acts
on the single bonds because it is shared between all closed bonds in
the cluster.

The forward rate is $g_i = (N_t-i) k_{on}(i)$ because each of the
$(N_t-i)$ open bonds can be the next to close. The single bond
association rate increases as the receptor-ligand distance $x_b(i)$
decreases and can be written as
\begin{equation} \label{kon}
k_{on}(i) = \frac{k_{on}}{Z} e^{-\frac{1}{2}k_b x_b(i)^2/\kt}
\end{equation}
where $Z$ is the partition function for a ligand in a harmonic potential
confined between $-\ell_b$ and $x_b(i)$, $k_{on}$ is the single bond
on-rate if receptor and ligand are in close proximity, and $x_b(i) =
F_b(i) / k_b$ follows from \eq{Force_i}. In our model, temperature $T$
essentially controls ligand mobility: the larger $T$, the broader the
ligand distribution. In the limit $T \to \infty$ (high ligand mobility,
$k_{on}(i) = k_{on}$) and $k_b \gg k_t$ (soft transducer, $F_b(i) = k_t
\ell / i$), our model simplifies to a case which has been used before to
study the stability of adhesion clusters under force \cite{uss:erdm04a}.
However, only the full model introduced here results in bistability.

\begin{figure}
\begin{center}
\includegraphics[width=0.8\columnwidth]{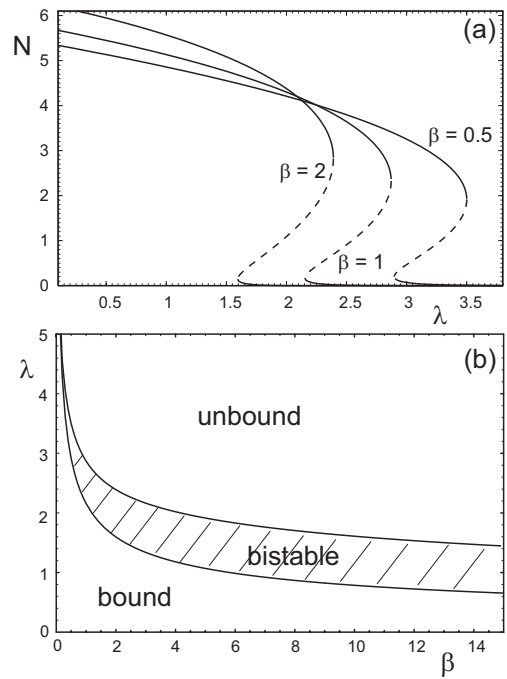}
\caption{(a) One-parameter bifurcation diagram showing stable (solid
  lines) and unstable (dashed lines) fixed points of the mean field
  equation \eq{DeterministicEquation} as function of unloaded
  receptor-ligand distance $\lambda = \ell/\ell_b$ and for three
  values of inverse ligand mobility, $\beta = k_b\ell_b^2/\kt = 0.5$,
  $1$ and $2$. The other parameters are $N_t=10$, $k_{on}/k_0 = 1$,
  $k_b / k_t = 1$ and $k_b \ell_b / F_0 = 0.1$. (b) Stability diagram
  constructed from the location $\lambda$ of upper and lower
  bifurcation points shown in (a) as function of $\beta$.}
\label{BifurcationDiagram}
\end{center}
\end{figure}

In order to demonstrate that our model leads to bistability, it is
sufficient to consider the mean number of closed bonds, $N(t) = \sum_{i
= 1}^{N_t} i p_i(t)$. Neglecting deviations from the mean, the time
evolution of $N(t)$ is described by the ordinary differential equation
\begin{equation} \label{DeterministicEquation}
\frac{dN}{dt} = g(N) - r(N)
\end{equation}
where reverse and forward rate are now interpreted as functions of a
continuous argument. The same mean field approach has been used before
to study the stability of adhesion clusters under force in the limits of
high ligand mobility and soft transducer, both for constant
\cite{c:bell78} and linearly rising force \cite{c:seif00}. In
\fig{BifurcationDiagram} we show the results of a bifurcation analysis
of \eq{DeterministicEquation}. In \fig{BifurcationDiagram}a, stable
(solid lines) and unstable (dashed lines) fixed points are shown as
function of the dimensionless unloaded receptor-ligand distance $\lambda
= \ell/\ell_b$ and for different values of the dimensionless inverse
ligand mobility $\beta = k_b\ell_b^2/\kt$. For small $\lambda$, a
single, stable fixed point exists at finite $N$. Here, the adhesion
cluster is bound because force on a closed bond is small and the density
of free ligands close to the receptors is large, therefore rupture
events are rare and can be balanced by rebinding. For large $\lambda$,
there is a single, stable fixed point at $N \approx 0$. Here, the
adhesion cluster is unbound because forces are large and ligand density
at the receptors is small, therefore rupture occurs frequently and
cannot be balanced by rebinding. The transition from bound to unbound
proceeds via a series of two saddle-node bifurcations, resulting in a
window of bistability. This bistability results from two positive
feedback mechanisms working in parallel. First, there is positive
feedback for rupture: as one bond breaks, force on the remaining bonds
is increased, thus increasing their dissociation rate. Second, there is
positive feedback for binding: as one ligand binds, receptor-ligand
distance is decreased and the binding rate for the other ligands
increases.  \fig{BifurcationDiagram}a also shows that with increasing
ligand mobility (decreasing $\beta$), the bistable region shifts to
larger $\lambda$ because the spatial distribution of ligands becomes
broader, thus stabilizing adhesion clusters through rebinding. In
\fig{BifurcationDiagram}b we construct a stability diagram by plotting
the $\lambda$ corresponding to the upper and lower bifurcation points as
a function of $\beta$. Again we observe that $\lambda$ decreases with
increasing $\beta$. For the lower bifurcation we find the scaling
$\lambda \sim \beta^{-1/2}$. For large $\beta$ (small ligand mobility),
the width of the two-state-region is fairly constant, while it decreases
for small $\beta$ (large ligand mobility) as the two curves converge.
 
\begin{figure}
\begin{center}
\includegraphics[width=0.8\columnwidth]{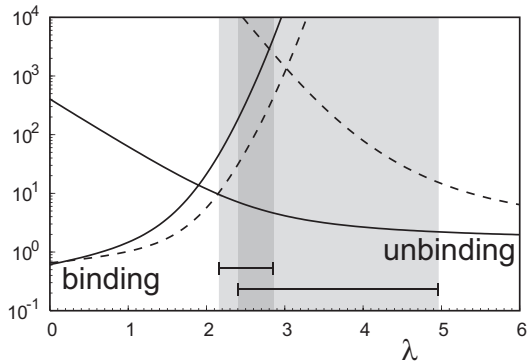}
\caption{Average binding and unbinding times in units of $k_0$ as
function of unloaded receptor-ligand distance $\lambda = \ell / \ell_b$
for $N_t=10$ (solid lines) and $25$ (dashed lines) at $\beta=1$. The
other parameters are as in \fig{BifurcationDiagram}.  The shaded areas
marked by bars are the windows of bistability (which overlap in the
darker region) from the mean field theory for $N_t = 10$ (left) and $N_t
= 25$ (right).}
\label{MeanFirstPassageTimes}
\end{center}
\end{figure}

To study the dynamics of adhesion clusters, the full stochastic model
of \eq{MasterEquation} has to be used. Bistability demonstrated in the
bifurcation analysis of the mean field theory now corresponds to
bimodality of the stationary probability distribution. Again,
coexistence of bound and unbound adhesion clusters is found for
intermediate values of the unloaded receptor-ligand distance
$\lambda$.  In the stochastic description, the system is able to
switch dynamically between these coexisting macrostates. Using
\eq{MasterEquation}, average binding and unbinding times can be
calculated as mean first passage times \cite{b:kamp92}. In
\fig{MeanFirstPassageTimes} we plot average binding and unbinding
times as function of the unloaded receptor-ligand distance $\lambda$
for cluster sizes $N_t = 10$ (solid lines) and $25$ (dashed lines). A
detailed analysis shows that their ratio effectively equals the ratio
of occupancy probabilities of the stationary solution of the master
equation. Therefore when binding and unbinding times are equal,
unbound and bound state are occupied with equal probability, thus
defining a stochastic transition point. The shaded areas in
\fig{MeanFirstPassageTimes} show the windows of bistability derived
from the mean field approach.  For $N_t=10$, the stochastic transition
point is below the deterministic range, while for $N_t=25$, it falls
into the deterministic range.  The definition of a range of
bistability is less appropriate in the stochastic case because here
the existence of a bimodal distribution is completented by information
on the transition times. For both deterministic and stochastic
descriptions, the $\lambda$-values for bistability increase with
$N_t$, so that for a given $\lambda$ the bound state becomes the only
stable macrostate when $N_t$ grows. In the stochastic treatment, this
leads to an super-exponentially fast increase of unbinding time with
$N_t$ for fixed $\lambda$, so that larger adhesions get effectively
trapped in the bound state and only small adhesions will show bistable
switching on an experimental time scale.

We finally discuss how our results apply to the integrin-fibronectin
bond. For the low affinity fibronectin-$\alpha_5\beta_1$-integrin bond
$k_0 \simeq 0.13\hz$ and $F_0 \simeq 10$ pN has been measured
\cite{c:li03} and we assume $k_{on}/k_0 \simeq 1$. Fibronectin is a
semi-flexible polymer with contour length $L \simeq 62\nm$,
persistence length $L_p \simeq 0.4\nm$ and equilibrium length $\ell_b
\simeq 11\nm$. Therefore the spring constant at small extensions can
be estimated to be $k_b = 3\kt / 2L_p L \simeq 0.25\pn\nm^{-1}$. From
this we get $k_b \ell_b / F_0 \simeq 0.27$ and $\beta \simeq 7.1$. In
order to estimate the effective spring constant of the force
transducer, we use the Hertz model for elastically deforming the cell
with a cylinder of radius $a$. Then $k_t = 2aE/(1-\nu^2) \simeq
0.27\pn\nm^{-1}$ where we have used $E \simeq 10\,\rm kPa$ for the
Young modulus, $\nu = 0.5$ for the Poisson ratio and $a \simeq 10\nm$
for the size of the adhesion cluster. Thus the ratio of force
constants is $k_b / k_t = 0.9$ and bistability for $N_t = 5$ is found
from the full stochastic model around $\lambda \simeq 0.75$.  The
actual unloaded cell-substrate distance is the unloaded
receptor-ligand separation $\ell$ plus the ligand resting length
$\ell_b$, that is $20\nm$. This indeed is the range of cell-substrate
distance at which mature cell-matrix adhesion starts \cite{c:cohe04}.

In summary, our model predicts that the early phase of cell-matrix
adhesion is characterized by rapid association and dissociation of small
adhesion sites which have attachment times in the range of several tens
of seconds. In the presence of appropriate signals, commitment to firm
adhesion can be induced through different mechanisms, including
recruitment of additional receptors and changes in binding affinity,
which increase the escape time from the bound macrostate. In this sense,
the structural mechanism for bistability described here resembles the
switch-like behavior often encountered in biochemical control
structures.

Acknowledgments: This work was supported by the Emmy Noether Program
of the German Research Foundation (DFG) and the Center for Modelling
and Simulation in the Biosciences (BIOMS) at Heidelberg.


\begin{thebibliography}{1}

\bibitem{c:geig02}
B.~Geiger and A.~Bershadsky.
\newblock Exploring the neighborhood: adhesion-coupled cell mechanosensors.
\newblock {\em Cell}, 110:139--142, 2002.

\bibitem{s:ange04}
D.~Angeli, J.~E. Ferrell, and E.~D. Sontag.
\newblock Detection of multistability, bifurcations, and hysteresis in a large
  class of biological positive-feedback systems.
\newblock {\em PNAS}, 101:1822--1827, 2005.

\bibitem{c:bell78}
G.~I. Bell.
\newblock Models for the specific adhesion of cells to cells.
\newblock {\em Science}, 200:618--627, 1978.

\bibitem{uss:erdm04a}
T.~Erdmann and U.~S. Schwarz.
\newblock Stability of adhesion clusters under constant force.
\newblock {\em Phys. Rev. Lett.}, 92:108102, 2004.

\bibitem{c:seif00}
U.~Seifert.
\newblock Rupture of multiple parallel molecular bonds under dynamic loading.
\newblock {\em Phys. Rev. Lett.}, 84:2750--2753, 2000.

\bibitem{b:kamp92}
N.~G. van Kampen.
\newblock {\em Stochastic processes in physics and chemistry}.
\newblock Elsevier, Amsterdam, 1992.

\bibitem{c:li03}
F.~Li, S.~D. Redick, H.~P. Erickson, and V.~T. Moy.
\newblock Force measurements of the $\alpha_5 \beta_1$ integrin-fibronectin
  interaction.
\newblock {\em Biophys. J.}, 84:1252--1262, 2003.

\bibitem{c:cohe04}
M.~Cohen, D.~Joester, B.~Geiger, and L.~Addadi.
\newblock Spatial and temporal sequence of events in cell adhesion: from
  molecular recognition to focal adhesion assembly.
\newblock {\em ChemBioChem}, 5:1393--1399, 2004.

\end{thebibliography}

\end{document}